\documentstyle[aaspp4,epsf]{article}

\begin{document}
\title{A new measure of $\sigma_8$ using the lensing dispersion in
high-$z$ type Ia SNe}
\author{Takashi Hamana\altaffilmark{1}\\
Yukawa Institute for Theoretical Physics, Kyoto University, Kyoto
606-8502, Japan\\
and\\
Toshifumi Futamase\\
Astronomical Institute, Tohoku University, Sendai
980-8578, Japan}
\altaffiltext{1}{Current address: Institut d'Astrophysique de Paris,
98bis Boulevard Arago, F-75014 Paris, France}
\begin{abstract}
The gravitational lensing magnification or demagnification due to
large-scale structures induces a scatter in peak magnitudes of high
redshift type Ia supernovae (SNe Ia).
The amplitude of the lensing dispersion strongly depends on that of
density fluctuations characterized by the $\sigma_8$ parameter.
Therefore the value of $\sigma_8$ is constrained by measuring the
dispersion in the peak magnitudes.
We examine how well SN Ia data will provide a constraint on the
value of $\sigma_8$ using a likelihood analysis method.
It is found that the number and quality of SN Ia data needed for
placing a useful constraint on $\sigma_8$ is attainable with Next
Generation Space Telescope.
\end{abstract}
\keywords{cosmology: observations --- cosmology: theory ---
gravitational lensing --- large-scale structure of universe ---
supernovae: general}

\section{Introduction}
It has long been recognized that the type Ia supernovae (SNe Ia) may
be a powerful tool for doing the cosmology for their homogeneity as
well as their very high luminosity.
For example, recent measurements of high-$z$ SNe Ia have provided 
useful constraints on values of the cosmological parameters, present
values of the density parameter $\Omega_m$ and normalized cosmological 
constant $\Omega_\Lambda$, though they have somewhat large confidence
interval in $\Omega_m$-$\Omega_\Lambda$ plane (Riess et al.~1998;
Perlmutter et al.~1999, hereafter SCP99). 
There are now two observational projects, the Supernova Cosmology
Projects\footnote{for more information on the Supernova Cosmology
Projects, see http://www-supernova.lbl.gov/.} and the High-Z Supernova 
Search\footnote{for more information on the High-Z Supernova Search,
see
http://cfa-www.harvard.edu/cfa/oir/Research/supernova/HighZ.html.}, 
involved in the systematic investigation of high-$z$ SNe Ia for
cosmological purposes.
With their great effort, it seems to be promising that the quantity as
well as the quality of the data are rapidly improved in near future.

The SNe Ia are, however, not perfect standard candles but have
scatter in their peak magnitudes.
There are mainly two sources of the scatter: 
One is due to the intrinsic heterogeneity in SNe Ia which has been
found empirically small, with a dispersion $\sigma_m\sim 0.3$mag in 
$B$ band (Branch 1998 and references cited therein). 
Moreover it has been also pointed out that using the observed correlations
between light-curve shape and luminosity in several different filters, 
the effective dispersion can be reduced to $0.12 \sim 0.17$mag
(Nugent et al.~1995; Hamuy et al.~1996; Riess, Press \& Kirshner 1996).
The another is the gravitational lensing magnification (or
demagnification) effect caused by the inhomogeneous distribution of
the matter between SNe Ia and us.

The lensing dispersion in the peak magnitudes due to large-scale
structures in the cold dark matter (CDM) models has been investigated
analytically (Frieman 1997; Nakamura 1997) and numerically
(Wambsganss et al.~1997; Wambsganss, Cen \& Ostriker 1998; Hamana,
Martel \& Futamase 1999). 
It has been found that the dispersion depends strongly on the
amplitude of fluctuations of the matter and their evolution, more
explicitly on $\sigma_8$, the rms fluctuation of the matter on
$8h^{-1}$Mpc, and on $\Omega_m$.
Furthermore, the lensing dispersion becomes larger than 0.1 at
redshift 0.5 if both $\Omega_m$ and $\sigma_8$ are larger than 1.
It should be noted here that compact virialized objects such like
individual galaxy or cluster of galaxies may also contribute to the
lensing magnification (Holz \& Wald 1998; Holz 1998).
However such nonlinear objects may cause a large magnification and
the existence of lensing galaxies could be confirmed, for example, by
deep imaging.
Even if a lensing galaxy is not discovered, such an exceedingly luminous
SN Ia should not be included in a normal SN Ia sample.
Therefore, in this paper, we do not take into consideration strong 
lensing effects as a source of the scatter.

The idea that the dispersion in peak magnitudes of the high-$z$ SNe Ia 
may be a probe of the amplitude of the density fluctuations was first
pointed out by Metcalf (1999).
He found that the amount and quality of data needed for placing useful
constraints on its value are attainable in a few years.
This method has the advantage over other methods such as the two-point
correlation functions of galaxies and cluster abundance 
in that the method is free from the unknown bias and the uncertain
luminosity-temperature relation in X-ray clusters of galaxies.
Metcalf (1999) parameterized the amplitude of the lensing dispersion
by one parameter $\eta_0$ which basically measures the amplitude of the
appropriately projected density fluctuations but can not determine the
values of $\Omega_m$ and $\sigma_8$ separately.
Since $\sigma_8$ is one of most important quantities to study the
evolution of the structures in the universe, it is worth exploring a
possibility of placing a meaningful constraint on its value using
the high-$z$ SN Ia data.

The purpose of this paper is to examine how well SN Ia data will
place constraints on the values of $\Omega_m$ and $\sigma_8$ in the
light of the rapid increase in discovery of SNe Ia with redshift
around or larger than 1 in near future.
For this purpose, we first re-examine the dispersion in the lensing
magnifications predicted using, so-called, the power
spectrum approach (Kaiser 1992; Nakamura 1997; Hamana et al.~1999) in
\S 2.
Special attention is paid to the scaling of the dispersion with
$\sigma_8$ and $\Omega_m$.
In \S3, we investigate a possible constraint on $\sigma_8$ that is
expected to be obtained from future SN Ia data using the likelihood 
analysis method, where the effect of $\sigma_8$ on the likelihood
function enters through the dispersion of peak magnitudes due to the
lensing magnifications.
We also show the contour map of the likelihood function in the 
$\Omega_m$-$\sigma_8$ plane calculated using the currently
available data in SCP99.
Although the current data does not provide a useful constraint on the
value of $\sigma_8$, that will be a help to see how does two parameter
degenerate in the plane.
General discussions including the possibility of observing the
SNe Ia at $z>1$ with large telescopes including the {\sl Next Generation
Space Telescope} (NGST) are given in \S4.

\section{Dispersion in gravitational lensing magnifications}
The variance of the lensing magnification, $\sigma_\mu^2$, of a point
like source such like SNe Ia due to the large-scale structures can be
estimated using both the Born approximation (Bernardeau, van Waerbeke
\& Mellier 1997; Schneider et al.~1998) 
and Limber's equation in Fourier space (Kaiser 1992; 1998).  
The variance is related to the density power spectrum, $P(k,w)$, by
(Nakamura 1997; Hamana et al.~1999),
\begin{equation}
\label{sigma_mu}
\sigma_\mu^2 (w) = 
{{9 \Omega_m^2} \over {2 \pi}} \left( {{H_0 \over c}} \right)^4 
\int_0^w dw' \left[{f_K(w') f_K(w-w') \over {f_K(w) a(w')}} \right]^2
\int_0^\infty dk kP(k,w').
\end{equation}
Here $w$ is the comoving radial distance, $a$ is the scale factor
defined by usual manner (e.g.~Weinberg 1972) and
normalized by its present value (i.e., $a_0=1$), and $f_K(w)$ is the
corresponding comoving angular diameter distance, defined as $K^{-1/2} 
\sin K^{-1/2} w$, $w$, $(-K)^{-1/2} \sinh (-K)^{-1/2} w$ for $K>0$,
$K=0$, $K<0$, respectively, where $K$ is the curvature which can be
expressed by $K=(\Omega_m+\Omega_\Lambda-1)H_0^2/c^2$.
It should be noticed here that assumptions on the linear evolution and 
Gaussianity of the density field have not been used in deriving the
equation (\ref{sigma_mu}).
We shall use the fitting formula of Peacock \& Dodds (1996) (PD96
hereafter) to describe the nonlinear evolution of density power
spectra. 
The relationship between the comoving distance and the
redshift $z$ (or equivalently the scale factor $a$) can be derived
from the Friedmann equation (Jain \& Seljak 1997): 
\begin{equation}
\label{w-z}
w (z) = {c \over {H_0}} \int_{1 /(1+z)}^1 da \left[\Omega_\Lambda a^4 
+(1-\Omega_m-\Omega_\Lambda) a^2 + \Omega_m a \right]^{-{1/2}}.
\end{equation}

In our recent paper, Hamana et al.~(1999) numerically investigated the
statistics of the weak gravitational lensing in CDM models performing
the ray-tracing experiments combined with P$^3$M $N$-body simulations.
We have compared the lensing dispersions obtained from the experiments
with the predictions of the analytical approach with the PD96's
fitting formula. 
We have found a good agreement between these two values within errors
caused by the force resolution in P$^3$M $N$-body simulations.
The analytical formula, eq.~(\ref{sigma_mu}), combined with PD96's
fitting formula is, therefore, a good approximation of the lensing
dispersion for the study presented in this paper.

\begin{figure}[t]
\begin{center}
\begin{minipage}{8cm}
\begin{center} 
\epsfxsize=8cm
\epsffile{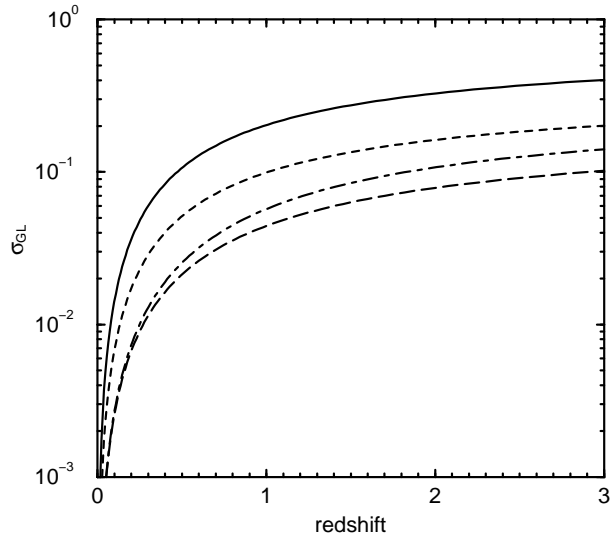}
\end{center}
\end{minipage} 
\caption[takashi_Fig1.eps]{$\sigma_{GL}$ versus source redshift.
The solid curves is for the model with
$(\Omega_m,\Omega_\Lambda,\sigma_8) = (1,0,1)$. the dashed line is for
$(1,0,0.5)$, the long dashed line is for $(0.3,0,1)$ and the long
dot-dashed line is for $(0.3,0.7,1)$.
The fitting formula of Peacock \& Dodds (1996) is used to describe the
nonlinear evolution of the density power spectra. \label{fig-1}}
\end{center}
\end{figure}

We consider CDM models. The transfer function, we adopted, is given by
Bardeen et al.~(1986). Throughout this paper, we take the Hubble
constant $H_0=70$km/sec/Mpc which is consistent with almost all of the
recent measurements (for a recent review, see Freedman 1999).
The dispersion in peak magnitude of SNe Ia due to lensing
magnification, $\sigma_{GL}$, relates to $\sigma_\mu$ by
$\sigma_{GL} \simeq 2.5\log(1+\sigma_\mu) \simeq 1.0857 \sigma_\mu$.
In figure \ref{fig-1}, we plot $\sigma_{GL}$ as a function of the
redshift for four different sets of parameters, $\Omega_m$,
$\Omega_\Lambda$ and $\sigma_8$.
It is clearly shown in Figure 1 that $\sigma_{GL}$ depends strongly on 
$\Omega_m$ and $\sigma_8$ but only weakly on $\Omega_\Lambda$, since
the effects of $\Omega_\Lambda$ on $\sigma_{GL}$ enters only though
the distance-redshift relation and the growth of the power spectrum.
In order to quantify the scaling of $\sigma_{GL}^2$ with $\sigma_8$ and
$\Omega_m$, we fit the dependence on these parameters to power laws.
Table \ref{table-1} provides such power-law fits for Einstein-de Sitter,
open and $\Omega_\Lambda$ dominated flat cosmologies.
It is evident from Table \ref{table-1} that $\sigma_8$ has a
comparable or a little weak dependence on $\sigma_{GL}^2$ compared 
with that of $\Omega_m$. 
Little deviation of the power of $\sigma_8$ from 2 (the relation,
$\sigma_{GL}^2 \propto \sigma_8^2$, is expected for a case of the
linear evolution of the density fluctuation spectrum)
is attributed to the effect of the nonlinear evolution of the density
fluctuations on small scales.

\begin{table}[h]
\caption{Scaling of $\sigma_{GL}^2$ with $\sigma_8$ and
$\Omega_m$.  \label{table-1}}
\begin{tabular}{cccc}
\hline
\hline
$z$ & $\sigma_8$ & $\Omega_m$ & $\Omega_\Lambda$\\
\hline
0.5 & ${\sigma_8}^{2.13}$ & 1 & 0 \\
0.5 & ${\sigma_8}^{2.30}$ & 0.3 & 0 \\
0.5 & ${\sigma_8}^{2.31}$ & 0.3 & 0.7\\
0.5 & 1 & ${\Omega_m}^{2.80}$ & 0 \\
0.5 & 1 & ${\Omega_m}^{2.48}$ & $1-\Omega_m$ \\
1 & ${\sigma_8}^{2.15}$ & 1 & 0 \\
1 & ${\sigma_8}^{2.31}$ & 0.3 & 0 \\
1 & ${\sigma_8}^{2.30}$ & 0.3 & 0.7\\
1 & 1 & ${\Omega_m}^{2.71}$ & 0 \\
1 & 1 & ${\Omega_m}^{2.26}$ & $1-\Omega_m$ \\
1.5 & ${\sigma_8}^{2.17}$ & 1 & 0 \\
1.5 & ${\sigma_8}^{2.32}$ & 0.3 & 0 \\
1.5 & ${\sigma_8}^{2.35}$ & 0.3 & 0.7\\
1.5 & 1 & ${\Omega_m}^{2.65}$ & 0 \\
1.5 & 1 & ${\Omega_m}^{2.14}$ & $1-\Omega_m$ \\
2 & ${\sigma_8}^{2.18}$ & 1 & 0 \\
2 & ${\sigma_8}^{2.32}$ & 0.3 & 0 \\
2 & ${\sigma_8}^{2.33}$ & 0.3 & 0.7\\
2 & 1 & ${\Omega_m}^{2.62}$ & 0 \\
2 & 1 & ${\Omega_m}^{2.07}$ & $1-\Omega_m$ \\
\hline
\end{tabular}
\end{table}

\section{The maximum likelihood analysis with the lensing dispersion}
Let us suppose that we observe $N$ SNe Ia having a peak magnitude
$m_i$ (corrected for $K$-correction, decline rate-luminosity relation,
dust extinction {\sl etc}) with a magnitude error $\sigma_{m,i}$ and 
redshift $z_i$.
The predicted magnitude-redshift relation is given by
\begin{equation}
\label{m-z}
m^{pred}(z)={\cal{M}}+5 \log {\cal{D}}_L(z,\Omega_m,\Omega_\lambda),
\end{equation}
where ${\cal{M}}$ is related to the peak absolute magnitude $M$ by
\begin{equation}
\label{calM}
{\cal{M}} = M + 5 \log \left({{c/H_0} \over {10\mbox{pc}}} \right),
\end{equation}
and ${\cal{D}}_L$ is the normalized luminosity distance defined by
\begin{equation}
\label{calD}
{\cal{D}}_L(z,\Omega_m,\Omega_\lambda) = {{H_0} \over c} (1+z) f_K(w(z)).
\end{equation}
In order to determine the parameters (in our case, $\Omega_m$,
$\Omega_\Lambda$, ${\cal{M}}$ and $\sigma_8$), we shall maximize the
Gaussian likelihood function defined by
\begin{equation}
\label{L}
{\cal{L}} = \prod_{i=1}^{N} { 1\over {\sqrt{2 \pi} \sigma_i}} \exp
\left( - {{(m_i - m_i^{pred})^2} \over {2 \sigma_i^2}} \right),
\end{equation}
where $\sigma_i^2 = \sigma_{m,i}^2+\sigma_{GL,i}^2$ in which we
have assumed that there is no correlation between the magnitude error
and that cased by the lensing magnification.

\begin{table}
\caption{Numerical values of $\sigma_{GL}$ and the numbers of SNe
Ia needed for $2\sigma$ detection of $\sigma_8$ for a case of
$\sigma_8=1$. In all cases, $\sigma_m$ is fixed to be
0.15. \label{table-2}}
\begin{tabular}{cccccccccc}
\hline
\hline
\multicolumn{2}{c}{Model} & \multicolumn{2}{c}{$z=0.5$} &
\multicolumn{2}{c}{$z=1$}  
& \multicolumn{2}{c}{$z=1.5$} & \multicolumn{2}{c}{$z=2$}\\
$\Omega_m$ & $\Omega_\Lambda$ & $\sigma_{GL}$ & $N$ & $\sigma_{GL}$ &
$N$ & $\sigma_{GL}$ & $N$ & $\sigma_{GL}$ & $N$  \\
\hline
1   & 0   & $1.06 \times 10^{-1}$ & 4.21 & $2.03 \times 10^{-1}$ &
4.14 & $2.74 \times 10^{-1}$ & 2.87 & $3.27 \times 10^{-1}$ & 2.45 \\
0.3 & 0   & $2.13 \times 10^{-2}$ & 3870 & $4.43 \times 10^{-2}$&  233
& $6.33 \times 10^{-2}$ & 65.0 & $7.87 \times 10^{-2}$ & 31.9 \\
0.3 & 0.7 & $2.54\times 10^{-2}$ & 1930 & $5.72 \times 10^{-2}$ & 93.8
& $8.47 \times 10^{-2}$ & 24.8 & $1.07 \times 10^{-1}$ & 13.0 \\
\hline
\end{tabular}
\end{table}

We now estimate, basically following Metcalf (1999), the number of SNe
Ia needed for a detection of the $\sigma_8$ value with a certain
significance level.
The precision with which a model parameter will be determined can be
estimated by ensemble average of the Fisher matrix. 
For the case of $\sigma_8$, that is given by
\begin{equation}
\label{sigma_sigma8}
\left[\sigma_{\sigma_8}^2\right]^{-1} 
= \left\langle - {{\partial^2 \ln {\cal{L}}} \over {\partial
{\sigma_8}^2}} \right\rangle 
= {1\over 2} \sum_{i=1}^N {{\left[ \partial \sigma_{GL,i}^2/\partial
\sigma_8 \right]^2} \over {( \sigma_{m,i}^2 + \sigma_{GL,i}^2 )^2}}.
\end{equation}
If we use the power-law fit for the lensing dispersion,
i.e.~$\sigma_{GL}^2 =
(\sigma_8/{\sigma_8^\ast})^{\gamma}\sigma_{GL}^{\ast 2}$,
where quantities with the asterisk refer to their values in a certain
model, moreover we assume that SNe Ia locate the same redshift and have the
same $\sigma_m$, then the above equation is simplified to 
$[\sigma_{\sigma_8}^2]^{-1} = N \gamma \sigma_8^{-2}
\sigma_{GL}^4 / 2 ( \sigma_{m}^2 + \sigma_{GL}^2)^2$.
In Table \ref{table-2}, we summarize required numbers of SNe Ia for
$2\sigma$ detection of $\sigma_8$ estimated under the above assumptions.
The redshift is taken to be $z=0.5$, 1, 1.5 and 2 with magnitude error
$\sigma_m=0.15$.
As Table {\ref{table-2} indicates, it is essential to observe SNe Ia
having a redshift larger than 1 for placing a meaningful constraint on
$\sigma_8$.
The above equation tells us that the number goes as $\sigma_m^4$ so
it is very sensitive to this parameter.
Reducing the magnitude error is, therefore, an another important point
of this method.
Similarly, $N\propto \sigma_8^{-2 \gamma}$, thus if the actual value of
$\sigma_8$ is small, for example 0.5, more than ten times the number
of SNe Ia compared with those in Table \ref{table-2} is needed for the 
detection with the same significance level.
It is, therefore, expected that this method will provide a strong
upper limit on the value of $\sigma_8$ rather than an actual
determination of the value. 

\begin{figure}[t]
\begin{center}
\begin{minipage}{8cm}
\begin{center} 
\epsfxsize=8cm
\epsffile{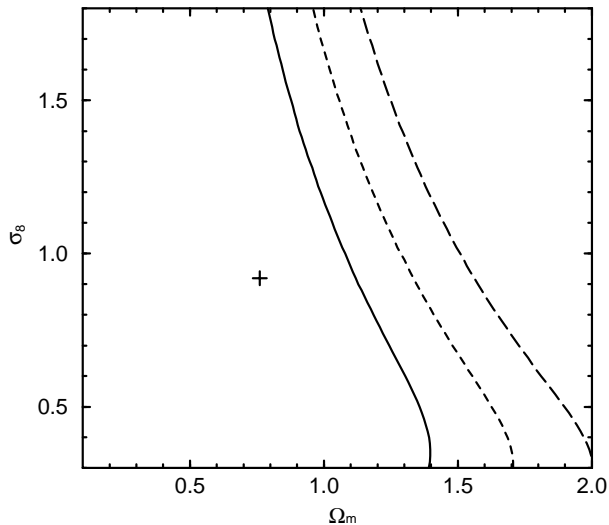}
\end{center}
\end{minipage} 
\caption[takashi_Fig2.eps]{Likelihood contours in the $\Omega_m$-$\sigma_8$ 
plane derived from the SN Ia data in Perlmutter et al.~(1999). 
The contours are plotted where $-2\ln{\cal{L}}/{\cal{L}}_{max}$ is
equal to $2.3$ (solid line), $6.2$ (dashed line) and $11.8$
(long-dashed line), corresponding approximately 1, 2 and 3$\sigma$
confidence contours for Gaussian likelihood function. The plus denotes
the position of the maximum likelihood. \label{fig-2}}
\end{center}
\end{figure}

Table \ref{table-2} also shows that the required number is very
sensitive to the cosmological model, especially to the density
parameter, because $\sigma_{GL}$ strongly depends not only on
$\sigma_8$ but also on $\Omega_m$ as shown in Table \ref{table-1}.
This immediately suggests that constraints obtained from SN Ia data
will degenerate in the $\Omega_m$-$\sigma_8$ plane.
We shall investigate this point using the currently available SN Ia
data of SCP99. 
We adopted SNe Ia used in 'primary fit' of SCP99 (their fit C).
We also adopted the corrected peak magnitudes and magnitude errors
summarized in Table 1 and 2 of SCP99, and thus we did not include the 
``stretch factor'' $\alpha$ (SCP99) in the light curve-luminosity
relation as a fitting parameter.
The likelihood function is computed in four-parameter space
($\Omega_m$, $\Omega_\Lambda$, ${\cal{M}}$ and $\sigma_8$).
In figure \ref{fig-2}, we plot the likelihood contours in the
$\Omega_m$-$\sigma_8$ plane, where we have not marginalized by
integrating the likelihood function over other parameters
($\Omega_\Lambda$ and ${\cal{M}}$) but have followed the peak, in
other words, we have not used the mean but the mode.
This does not make any significant difference as we will show below.
In the lower-left region in Figure \ref{fig-2} where $\Omega_m$ and
$\sigma_8$ are small, no useful constraint is provided.
This limitation comes from the fact that $\sigma_{GL}$ is smaller
than $\sigma_m$ for the models with a small $\Omega_m$ and $\sigma_8$.
Therefore it will be the case even if we have a large, very high-$z$
$(z>1)$ SN Ia sample.
This limitation can be improved only by reducing $\sigma_m$.
On the other hand, the upper-right region of Figure \ref{fig-2} is
relatively well constrained.
One may find in Figure \ref{fig-2} that the slope of the contour lines
in the $\Omega_m$-$\sigma_8$ plane are steeper than $-0.5$.
The reason for this is that the dependence of $\sigma_{GL}$ on
$\Omega_m$ is stronger than that on $\sigma_8$ as was shown in
Table \ref{table-1}, and the effect of $\Omega_m$ on the likelihood
functions also enters through the magnitude-redshift relation.
It may be, therefore, said that SN Ia data will hardly place a lower
limit on the value of $\sigma_8$, but a future large, very high-$z$ SN
Ia sample can provide a useful upper limit on the value of
$\sigma_8$. 

\begin{figure}[t]
\begin{center}
\begin{minipage}{8cm}
\begin{center} 
\epsfxsize=8cm
\epsffile{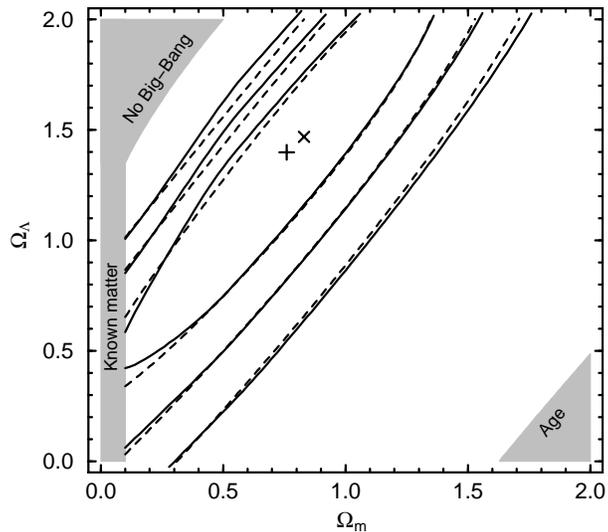}
\end{center}
\end{minipage}
\caption[takashi_Fig3.eps]{Likelihood contours in the 
$\Omega_m$-$\Omega_\Lambda$ 
plane derived from the SN Ia data in Perlmutter et al.~(1999).
The solid lines (dashed liens) and the plus (cross) are for the model
with (without) taking the lensing dispersion into consideration. 
The contours are plotted where $-2\ln{\cal{L}}/{\cal{L}}_{max}$ is
equal to $2.3$, $6.2$ and $11.8$.
The cross and plus denotes the positions of the maximum likelihoods. 
The shaded regions are ruled out by other constraints: The ``no
big-bang'' region at upper left, $\Omega_m<0.1$ is inconsistent with 
the amount of matter observed and we have simply taken $H_0 t_0>0.6$
for age. \label{fig-3}}
\end{center}
\end{figure}

One may question whether the lensing dispersion has any influence
on the likelihood contours in the $\Omega_m$-$\Omega_\Lambda$ plane.
In Figure \ref{fig-3}, we plot the likelihood contours calculated with 
and without taking the lensing dispersion into account.
The likelihood contours for the model without lensing dispersion are
identical to those of SCP99 (their fit C).
Figure \ref{fig-3} clearly indicates that the lensing dispersion has no
significant effect on the constraints on $\Omega_m$ and
$\Omega_\Lambda$, because the effect of the lensing dispersion on the
likelihood function enters only through the dispersion. 
Therefore, the conclusion of SCP99 and also that of Riess et al.~(1998) are
not changed by the lensing dispersion due to large-scale structures. 

\section{Discussion}
It may seem that the method proposed in this paper is not useful
compared with, e.g., the cluster abundance which have provided a tight
limit on the value of a combination of $\Omega_m$ and $\sigma_8$
(Eke, Cole \& Frenk, 1996; Kitayama \& Suto 1997).
However one should remember that the theoretical prediction of the
cluster abundance involves some uncertainties such like the
X-ray luminosity-temperature relation and the bias.
Our method is completely independent of the other methods in the
sense that it is free from the relation between the distribution of
dark matter and that of luminous matter, it can be a direct measure of
$\sigma_8$.
The combined study of these methods will provide a reliable constraint 
in the $\Omega_m$-$\sigma_8$ plane.

The most important point in using the SNe Ia as a probe of $\sigma_8$ 
is, of course, to observe them at higher redshift.
So far, there is no detection of SN Ia at $z>1$.
Gilliland, Nugent \& Phillips (1999) detected a likely SN event in a
revisit to Hubble Deep Field, it was associated with the galaxy at
$z=1.32$ (photometric), but no confirming spectrum of the SN was
obtained.
As this indicates, the main difficulty will be spectroscopy of SNe.
The region of spectrum that is used to do the light-curve
correction redshifts to the infrared.
The peak magnitude of a SN Ia at $z=1.5$ is expected to be $m_I\sim26$
(Gilliland et al.~1999; Dahl\'en \& Fransson 1999).
Direct spectroscopy will be very difficult for existing 8-10m
telescope below 25th magnitude, but it will be possible with
NGST\footnote{for more information on the Next Generation Space
Telescope see http://ngst.gsfc.nasa.gov/.}. 
A precise prediction of the number of SNe Ia at very high-$z$ is a
difficult task due to uncertainties in the cosmic star formation rate
and the progenitor's life time.
Dahl\'en \& Fransson (1999) made a prediction of $75 \sim 400$ SNe Ia per 
square degree down to $I_{AB}=27$ whose typical redshift will be
$z\sim1$ and have a broad redshift distribution to $z\sim2$.
They also predicted that $5 \sim 25$ SNe Ia will be detectable per NGST
field down to $K'=31.4$.
Therefore the number and quality of SN Ia data needed for for placing
a useful constraint on $\sigma_8$ is attainable with NGST.

We have not considered the possible evolution of SNe Ia properties or
galactic environments which are of great concern for using
the SNe Ia for cosmological purposes.
If the intrinsic dispersion of the peak magnitude increases with
redshift, the number of SNe Ia needed for placing a meaningful
constraint increases rapidly.
The systematic error in the peak magnitude provides incorrect
constraints not only on $\Omega_m$ and $\Omega_\Lambda$ but also on
$\sigma_8$ because these parameters are mutually related so that they
have to be determined simultaneously.
The detailed study of the possible evolutions will, of course, be a key
to obtain the correct constraints on these parameters.
The quantitative study of these issues will be done in elsewhere.

\acknowledgments
We would like to thank an anonymous referee for a careful report that
helped to improve this paper.
T.H.~is grateful to IAP where this work has been completed.
T.H.~acknowledges a C.O.E.~postdoctoral fellowship at Yukawa Institute
for Theoretical Physics, Kyoto University.
Numerical computation in this work was carried out at the 
Yukawa Institute Computer Facility.


\end{document}